# Self – assembly of model surfactants as reverse micelles in nonpolar solvents and their role as interfacial tension modifiers

E. Mayoral[1], J. A. Arcos-Casarrubias[2] and A. Gama Goicochea[2*]

[1]Instituto Nacional de Investigaciones Nucleares, Carretera México Toluca, La Marquesa Ocoyoacac, Estado de México, Mexico

[2]Posgrado en Ingeniería Química, División de Ingeniería Química y Bioquímica, Tecnológico de Estudios Superiores de Ecatepec, Av. Tecnológico, 55210, Estado de México, Mexico

## Abstract

The self-assembly of linear surfactants into reverse micelles (RMs) in nonpolar solvents, and their efficiency in reducing the interfacial tension is studied using dissipative particle dynamic simulations. Given the importance of RMs as thickeners, among many other applications, their properties are studied here when formed in oil and in supercritical carbon dioxide (scCO$_2$). Our simulations are found to be in agreement with experimental results of surfactant self-assembly in scCO$_2$ that found viscosity increments of up to 90% with 10 wt % of surfactants. The role played by a small number of water molecules in RM formation is studied as well in both solvents, corroborating experiments reporting the enhancement of RM formation with the addition of a small quantity of water. The dynamics of water surfactant aggregation in nonpolar solvents is also discussed. Lastly, the performance of the model surfactant as an interfacial tension modifier is studied in detail. The results show that RMs are more easily formed in oil than in scCO$_2$ and that the effectiveness of the surfactant in reducing the interfacial tension lies in its preference to adsorb at interfaces rather than self – association.

---

[*] Corresponding author. Electronic mail: agama@alumni.stanford.edu

# I INTRODUCTION

Micelles are stable self-assembled structures of amphiphilic molecules that have a polar head and a nonpolar tail, such as surfactants and lipids. Molecular self-assembly is promoted by the relationship between various interactions like van der Waals, hydrophobic-hydrophilic interactions, π−π stacking, hydrogen bond and electrostatics forces [1-3]. As a function of the spatial configuration and chemical characteristics of the molecules participating, diverse categories of micellar structures arise, such as bilayers, micelles, and reverse micelles [4]. In water or polar solvents, micelles have their hydrophobic ends inside forming a core and the hydrophilic shells outside [5, 6]. Opposed to micelles the phenomenon of self-assembly of amphiphilic molecules in nonpolar solvents gives rise to the emergence of aggregates known as reverse (or inverted) micelles (RMs) [4]. RMs are characterized by having the polar head of the surfactants at the core of the aggregate, and the nonpolar tails extending into the solvent. RMs can emerge in nonpolar solvents without water [7−13], but it is known that the presence of water influences the micellization [14−16]. RMs have various applications in industrial and basic research fields such as, extraction processes [17], emulsion polymerization reactions [18], synthesis of nanomaterials [19], drug delivery [20] and to study confinement effects and dehydration on biological molecules [21], among others. Lately, one of their most important applications are as thickeners of supercritical carbon dioxide (sc$CO_2$) for enhanced oil recovery and fracking [17, 22 – 24]. As an environmentally friendly solvent, sc$CO_2$ is nontoxic, nonflammable, easily recyclable and inexpensive [23]. The majority of the industrially used surfactant molecules are inefficient in producing stable RMs in sc$CO_2$ due to their small solubility [25]. Improving the stability of RMs in supercritical $CO_2$ would allow their large-scale application in enhanced oil recovery.

The first example of $CO_2$ viscosity thickener using anisotropic reversed micelles was reported by Trickett et al. [23]. They presented a methodology to thicken dense liquid $CO_2$ using self-assembly to produce $CO_2$ compatible fluorinated di-chain surfactants. Their conclusions showed that it is feasible to regulate surfactant self-assembly to produce long, narrow reversed micellar rods in dense $CO_2$ capable to enhance its viscosity by up to 90% while adding 10 wt % of surfactants [23]. DeSimone and co-workers [26, 27] developed and studied surfactants with polymeric heads and $CO_2$-philic perfluoro octyl acrylate tails. These surfactants produce a polymeric nucleus where the emulsion polymerization can be performed. Another category of surfactants with two tails, one alkane and the other one perfluoro alkane, known as di-chain or hybrid surfactants, have been shown to produce stable RMs with aqueous cores in $sCO_2$ [28 – 30]. Schurtenberger and collaborators [31] reported an increase in the viscosity of reverse micellar solutions of lecithin in isooctane by up to a factor of $10^6$ upon the addition of a small quantity of water. This trend can be detected in a significant range of distinct organic solvents where the addition of water stimulates the growth of macroscopically isotropic and thermodynamically stable viscous macromolecular solutions [31].

To acquire more knowledge about RMs formation, it is educational to develop a molecular – level understanding using, for example, molecular simulation. Different molecular simulation approaches have been used to study the morphology, dynamics, and rheology of surfactant aggregates in polar and nonpolar solvents [32 – 46]. Their predictions help analyze many important properties related with the micellization processes such as size, shape, surface roughness, and the internal structure of micelles. In fact, these studies have assisted in explaining previous experimental results [32,40]. Yet not many simulation studies have explored the phenomenon of surfactant self-assembly in nonpolar solvents, which results in the formation of RMs [47 – 49]. Salaniwal and collaborators [30] were the first to study the formation of stable RMs in scCO$_2$ using molecular simulations. They reported the earliest molecular simulations of the stable self-assembly of spherical aggregates of di-chain surfactants in scCO$_2$ [30], corresponding with earlier experimental reports by Eastoe and co-workers [28], and Johnston and co-workers [29]. Their results revealed that the shape and size of the RMs depend on the water-to-surfactant ratio. Senapati et al. [50, 51] reported molecular dynamics simulations of the RMs corresponding with fluorinated and hydrogenated surfactants in water/CO$_2$ systems. Lu and Berkowitz [52] reported simulations of RMs in a three-component system containing fluorinated surfactant, water and scCO$_2$ as a solvent. The effect of temperature and surfactant tail length has been reported also [53]. The need to understand the microscopic mechanisms that play a role in more complex systems at different concentrations and thermodynamic conditions using shorter simulation times, calls for accurate mesoscale techniques. It has been shown recently that a multiscale simulation based on atomistic and mesoscopic scales of polymers' role in thickening CO$_2$ offers useful insights into their aggregation mechanisms [54]. Using this approach, the role of a small amount of water in the system configuration was also examined [54].

In this work, we report a numerical simulation study for the self-assembly of inverse micelles and the performance as an interfacial tension modifier of a model linear surfactant. The interaction model used is the mesoscale dissipative particle dynamics approach (DPD) [55 – 57]. This work is divided in five main sections. After this Introduction, the section Models and Methods presents the interaction models, surfactant structure and solvents used in this work. Simulation details are included in section three and in section four the results and their discussion are presented in two parts. In the first one we presented the self-assembly of RMs of the model surfactant with and without water in two nonpolar solvents, oil (a strong nonpolar solvent) and scCO$_2$ (poor nonpolar solvent). The formation of RMs is analyzed and the effect of the water added is discussed. The dynamics of aggregation is also presented. In the second part, the performance of this surfactant as an interfacial tension modifier is studied. The change in the interfacial tension is analyzed as a function of the concentration of surfactant and the effectiveness of this surfactant is discussed as a function of its potential of mean force. Lastly, section five includes the main conclusions. This work provides answers about the association mechanisms of the surfactants forming reverse micelles in connection with the effectiveness of the surfactants in reducing the interfacial tension. Additionally, we show how the properties of the solvent affect the formation of reverse micelles. These results are useful for the design of surfactants that can help increase the viscosity of solvents, such as supercritical $CO_2$, for applications in enhanced oil recovery and fracking, among other uses.

## II MODELS AND METHODS

The fundamental structure of the DPD simulation method is essentially equivalent to the standard molecular dynamics' procedure [55 – 58]. One important difference is that in DPD the particles whose motion is solved are coarse-grained virtual entities representing molecular or atomic clusters known as beads. Each DPD particle has a momentum and position which is calculated by solving Newton's second law of motion at finite time steps, making use of the total force acting on it by pairs. The simple interparticle interactions of the DPD model make it possible to integrate the equations of motion with time steps of the order of picoseconds [55 – 57]. This is about three orders of magnitude greater than the time step used generally in atomistic simulations [58]. The total force acting on any two particles $i$ and $j$ is the sum of three terms: a conservative force ($\boldsymbol{F}_{ij}^{C}$), a dissipative force ($\boldsymbol{F}_{ij}^{D}$), and a random ($\boldsymbol{F}_{ij}^{R}$) force. The net force acting on the $i$-th particle, $\boldsymbol{F}_i$ is given by:

$$\boldsymbol{F}_i = \sum_{i \neq j} \boldsymbol{F}_{ij}^C + \boldsymbol{F}_{ij}^D + \boldsymbol{F}_{ij}^R \ , \tag{1}$$

where the DPD interparticle force exerted by the particle $i$ on the particle $j$ is additive in pairs. The dissipative $\boldsymbol{F}_{ij}^D$ and the random forces $\boldsymbol{F}_{ij}^R$ are defined as $F_{ij}^D = -\gamma \omega^D(r_{ij})[\hat{\boldsymbol{r}}_{ij} \cdot \boldsymbol{v}_{ij}]\hat{\boldsymbol{r}}_{ij}$ and $F_{ij}^R = \sigma \omega^R(r_{ij})\xi \hat{\boldsymbol{r}}_{ij}$ respectively; $\sigma$ is the noise amplitude and $\gamma$ is the friction coefficient, $\boldsymbol{r}_{ij} = \boldsymbol{r}_i - \boldsymbol{r}_j$, $r_{ij} = |\boldsymbol{r}_{ij}|$ and $\hat{\mathbf{r}}_{ij} = \mathbf{r}_{ij}/\mathrm{r}_{ij}$. Here $k_B$ is Boltzmann's constant, $\xi_{ij} = \xi_{ji}$ are random numbers between 0 and 1 with Gaussian distribution and unit variance and $\boldsymbol{v}_{ij} = \boldsymbol{v}_i - \boldsymbol{v}_j$ is the relative velocity between beads. To ensure that a Boltzmann distribution is reached at equilibrium $\sigma$ and $\gamma$ are related by $k_BT=\sigma^2/2\gamma$ due to the fluctuation−dissipation theorem [56]. The factors $\omega_D$ and $\omega_R$ are weight functions that depend on distance and become to zero for $r \geq r_c$ where $r_c$ represents a cutoff radius. This cutoff distance $r_c$ is selected as the reduced unit of length $r_c = 1$ and is the intrinsic length scale of the DPD model. For the conservative force $\boldsymbol{F}^C_{ij}$, a smooth, a linearly decaying repulsive interaction is used between each particle pair expressed as [57]:

$$F_{ij}^C = \begin{cases} a_{ij}\left(1 - \frac{r_{ij}}{r_c}\right)\hat{r}_{ij}, & r_{ij} < r_c \\ 0, & r_{ij} \geq r_c \end{cases} \tag{2}$$

where, $a_{ij}$ is the repulsive DPD parameter acting between the pair of particles. The masses of all particles are equal, with reduced unit of mass $m_i = 1$. To model the surfactants, DPD beads are connected by harmonic forces given by:

$$\boldsymbol{F}_{spring}(r_{ij}) = -k_0(r_{ij} - r_0)\hat{\boldsymbol{r}}_{ij}. \tag{3}$$

In Eq. (3) $k_0$ is the spring constant and $r_o$ the equilibrium distance of the spring. The interfacial tension (IFT) $\gamma^*$ is calculated from the components of the pressure tensor, as follows:

$$\gamma^* = L_z^*\left[\langle P_{zz}^*\rangle - \frac{1}{2}\left(\langle P_{xx}^*\rangle + \langle P_{yy}^*\rangle\right)\right], \tag{4}$$

where $\langle \cdots \rangle$ is the time average of the components of the pressure ($P_{ii}^*$) tensor and $L_z^*$ is the length of the simulation cell along the $z$ – direction. The asterisks indicate that expressions are given in dimensionless units. The components of the pressure tensor are obtained from the virial theorem [58].

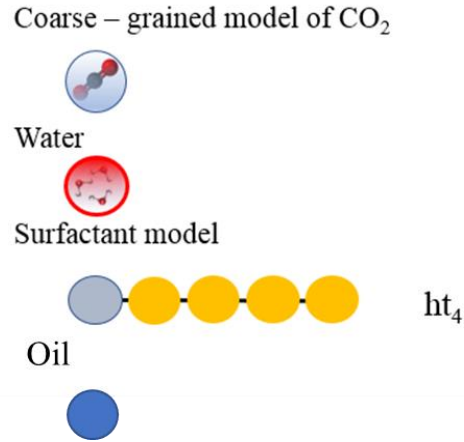

**Figure 1.** Components of the systems modeled in this work. The model surfactant is called ht$_4$ because it has a one head bead and 4 beads in its tail. Water is modeled as a single DPD bead, with coarse – graining degree equal to three. CO$_2$ and oil are also modeled as a single bead (in blue).

In this work, for the analysis of RM formation a three-component system is modeled, constituted by a linear surfactant called ht$_4$ with a polar head (h) and a four – bead nonpolar tail (t), see Fig. 1. The second component is water, which is added in small amounts and is represented by a single DPD bead. These two components are immersed on a nonpolar solvent that can be oil (model 1) or scCO$_2$ (model 2) represented by the monomeric structures shown in Fig. 1.

## III SIMULATION DETAILS

All the simulations reported here were performed in reduced DPD units under constant concentration, volume, and temperature. The time step chosen to integrate the equation of motion was $\Delta t^* = 0.03$; the volume of the cubic simulation box for the inverse micelle cases, both in oil and in CO₂ is $L_x \times L_y \times L_z = (17.1r_C)^3$. For the calculation of the water – oil interfacial tension as a function of surfactant concentration, the dimensions of the rectangular parallelepiped box are $L_x \times L_y \times L_z = 17.5 \times 17.5 \times 50 r_C^3$. For the inverse micelle cases there are $1.5 \times 10^4$ DPD beads in the simulation box, while for the water – oil interfacial tension system there are $4.6 \times 10^4$ DPD beads. These choices yield a global number density equal to 3. This is done so that the DPD equation of state remains invariant with respect to changes in the values of the interaction parameter of the conservative DPD force [59]. Periodic boundary conditions are applied in all faces of the simulation box. The values of the constants involved in the dissipative and random forces are chosen as $\sigma = 3$ and $\gamma = 4.5$, respectively, so that $k_B T^* = 1$. The cut-off radius is $r_c = 1$, and the masses are $m_i = m = 1$. The surfactants ht₄ modeled here are linear chains of beads linked by springs with the spring constant $k_0 = 100(k_B T/r^2{}_c)$ and $r_0 = 0.7 r_c$ [55, 59]; see Eq. (3). The coarse graining degree chosen here is three. This choice, along with the reduced value of the isothermal compressibility of water at room temperature, yield the value $a_{ii} = 78.3$ for the non – bonding interactions between particle of the same type [55,57], see Eq. (2). The full matrix of interaction parameters that define the maximum values of the non – bonding conservative DPD force between the various beads make up our systems, $a_{ij}$, is displayed in Table 1 for strong nonpolar solvent as oil as a solvent (model 1). In Table 2 one finds the interactions for scCO₂ (model 2), as a nonpolar solvent.

**Table 1.** DPD conservative interaction parameters used in model 1 when the strong nonpolar fluid is monomeric oil beads with water and the ht₄ surfactant.

| $a_{ij}$ | Water | Oil | Head | Tail |
|---|---|---|---|---|
| **Water** | 78.3 | 140.0 | 50 | 140.0 |
| **Oil** | | 78.3 | 140.0 | 78.3 |
| **Head** | | | 90.0 | 140.0 |
| **Tail** | | | | 78.3 |

**Table 2.** DPD conservative interaction parameters used in model 2 for scCO$_2$ as the nonpolar fluid, with water and ht$_4$ surfactants.

| $a_{ij}$ | Water | CO$_2$ | Head | Tail |
|---|---|---|---|---|
| **Water** | 78.3 | 103.61 | 50 | 140.0 |
| **CO$_2$** | | 78.3 | 103.61 | 78.3 |
| **Head** | | | 90.0 | 140.0 |
| **Tail** | | | | 78.3 |

For the case of oil (Table 1) the interactions were chosen heuristically so that the water – oil repulsion was strong, while hydrophilic – like interactions (water – head) were attractive. The interaction between surfactants' heads was chosen more repulsive than those between particles of the same type ($a_{ii} = 78.3$ in Table 1) to mimic the repulsion of the heads in ionic surfactants [46]. As for the case of model 2 in Table 2, these values were chosen so that the interfacial tension between water and scCO$_2$ at 313.15 K and 20 MPa is close to 25 mN/m [60]. Water and surfactant beads, whose relative distance was $r \leq 3r_C$ [53], were considered as belonging to an aggregate. All simulations were run for at least 50 blocks of 2×10$^4$ time steps, with the first 25 blocks used for equilibration and the rest for the production phase. At the beginning of the simulations all the components were distributed randomly and during the equilibrium phase surfactants formed self-assembled aggregates or migrate to the interfaces. When the surfactant concentration was increased, the number of solvent beads was reduced equally so that the global number density remains constant.

## IV RESULTS AND DISCUSSION

IV.1 Self-assembly of inverse micelles of $ht_4$ in oil and in $scCO_2$.

We begin by studying the self-assembly of $ht_4$ surfactants in oil without water, starting with a random configuration of 100 $ht_4$ molecules dissolved in oil. Figure 2(a) shows the surfactants' self-association in oil (which is omitted in the figure for clarity). Due to the high repulsion between the head of $ht_4$ and the solvent, surfactant molecules cluster into few small micelle-like aggregates. The tail of $ht_4$ stretches out into the surrounding oil because of its affinity with the solvent. Most of the amphiphilic molecules remain unassociated in the solvent.

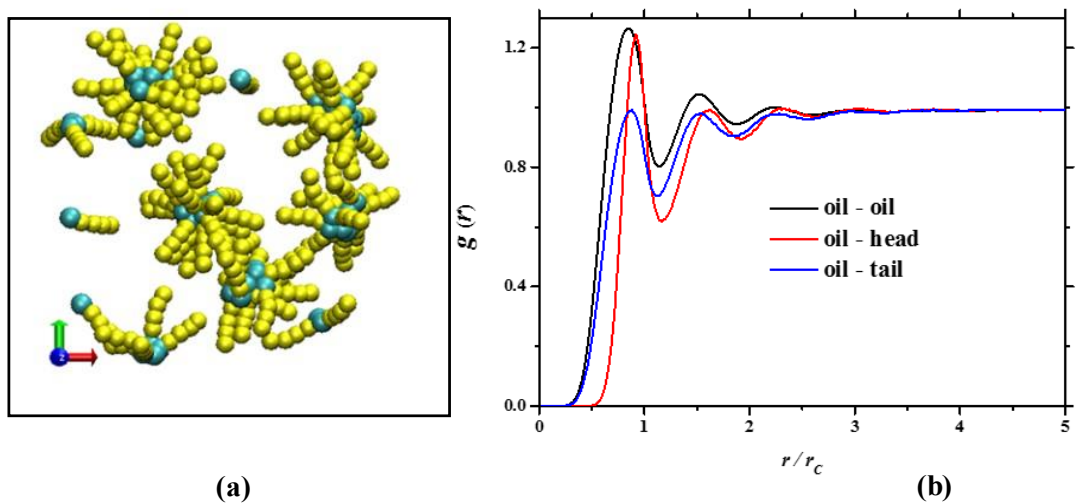

(a)                                                       (b)

**Figure 2**. **(a)** (Color online) Snapshot of the $ht_4$ surfactants self-association in oil, which is not shown for clarity. The beads that make up the surfactants' heads are shown in cyan and the tail beads are shown in yellow. There are 100 surfactant chains. **(b)** (Color online) Radial distribution functions between the $ht_4$ surfactants' heads and tails, and oil.

Figure 2(b) presents the spatial correlations from radial distribution function (RDF) for oil-oil, oil-head and oil-tail components of the system shown in Fig. 2(a). The strongest correlation is between solvent beads (oil-oil, in black solid line in Fig. 2(b)). The RDF for oil-tail is the closest to the one for oil with itself (blue line in Fig. 2(b)), because of the affinity of their interactions, see Table 1. Then solubility of $ht_4$ in oil is driven by the strong correlation between the nonpolar part (tail) of the surfactant and the solvent. The RDF for head-oil shows weak correlation with the solvent (read line in Fig. 2(b)), as expected. This nonaffinity explains why some RMs form in the system by the association between the heads of the surfactant, so that they avoid interacting with the solvent.

In most of the processes where surfactants have key applications, the presence of water is practically inevitable, though it is found at low concentrations. One of the effects of water present in small quantities in nonpolar solvents with surfactants is to promote the formation of RMs [23,31]. For this reason, we investigate the effect of water in the self-assembly mechanism of our system and proceed to add water molecules to the oil – $ht_4$ system. We start with a random distribution of 100 $ht_4$ molecules and 100 water molecules in oil. As shown in Fig. 3, water molecules attract each other and the heads of the surfactants, leading to the formation of RMs.

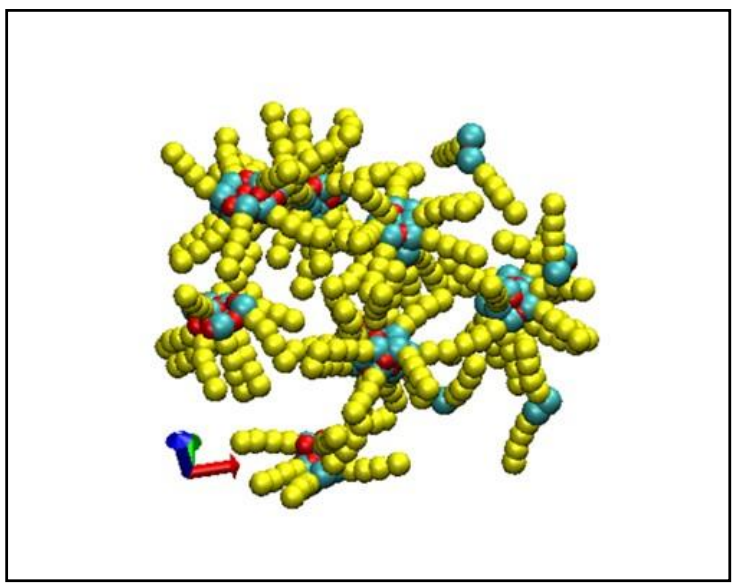

**Figure 3**. (Color online) Snapshot of 100 water molecules (red) with 100 $ht_4$ surfactant chains, in oil (not shown for clarity). The $ht_4$ heads are shown in cyan and the tails in yellow.

Figure 3 shows the formation of surfactant aggregates in the system that have the characteristics of RMs. These are self-assembling structures presenting aqueous nuclei surrounded by the ht$_4$ molecules, whose polar headgroups are submerged in the nucleus while the nonpolar tails form a crown. Figure 4 shows the density profiles for the surfactant/water/oil system at two different concentrations of surfactants (1.7 wt % and 2 wt %). It is found that the head and water distributions follow closely one another, corresponding with their distribution in the solvent forming the core of the RMs.

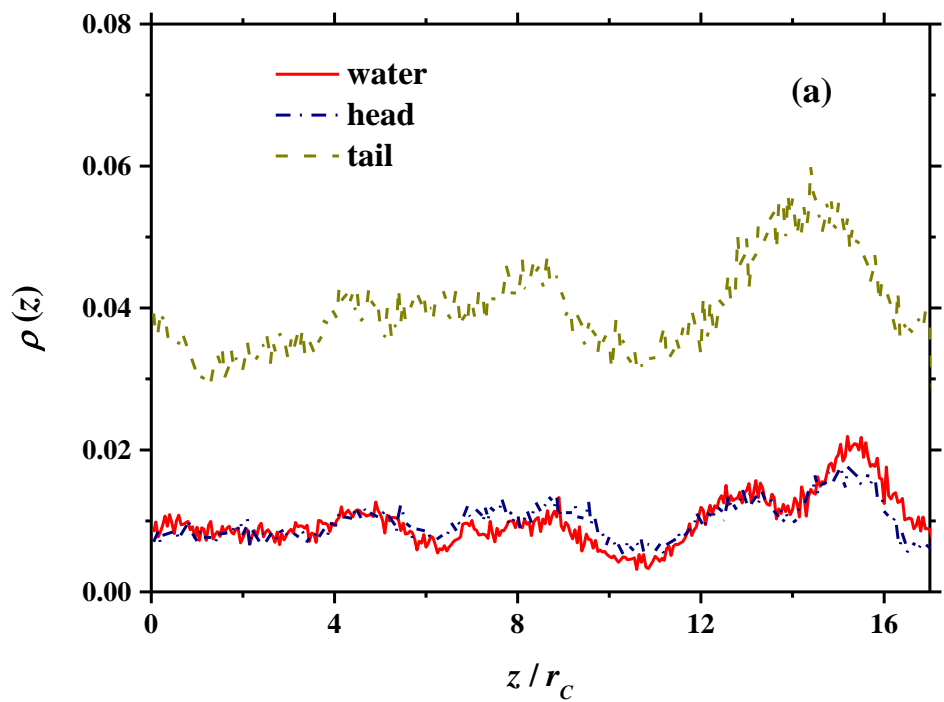

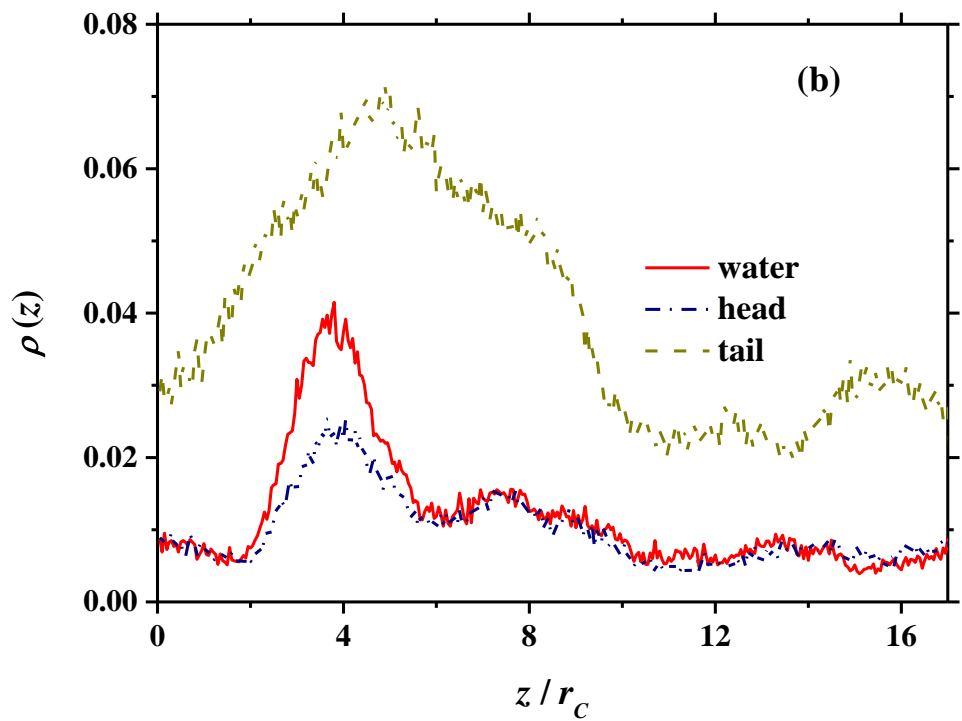

**Figure 4**. (Color online) Density profiles of the ht$_4$ surfactants' heads (dash – dotted line, navy blue), which follow closely the water beads profile (red), forming reverse micelles. The tails' profiles are shown by the dashed line in dark yellow. Two surfactant concentrations are shown: 1.7 wt % (a) and 2 wt % (b). The oil density profile is omitted, for clarity.

Having recognized that our mesoscopic model of a surfactant molecule does display the growth of stable reverse micelle – like structures in oil, we proceed to explore the structural characteristics of these aggregates. Figure 5 shows the RFDs for the water - ht$_4$ – oil system. The strongest correlation is between the head beads and the water beads forming clusters (Fig. 5(a)). This is indicative of the formation of stable self-organized structures driving forming RMs with the core of these aggregates made up of water and head groups. The water-oil and water-tail RDFs present weaker correlations, as is expected due the low affinity between the hydrophobic part of the surfactant and the nonpolar solvent with water (Fig. 5(a)). The water-tail RDF shows that the tails form the corona enveloping the nuclei, because their correlations are displaced to larger relative distances. The spatial correlations with the solvent are presented in Fig. 5(b), where one finds that the RDF for oil-tail is the closest to the one for oil with itself, promoting the solubility of ht$_4$ in the solvent. The micellar structures are distributed uniformly into the solvent. The mechanism that drives the formation of RMs in oil is the strong association of water with the surfactants' heads (solid red line, in Fig. 5(a)), *and* at the same time, the strong correlation of oil with the surfactants' tails (dotted blue line, in Fig. 5(b)). This is the same mechanism that becomes operative when modeling scCO$_2$ instead of oil, as shall be shown in what follows.

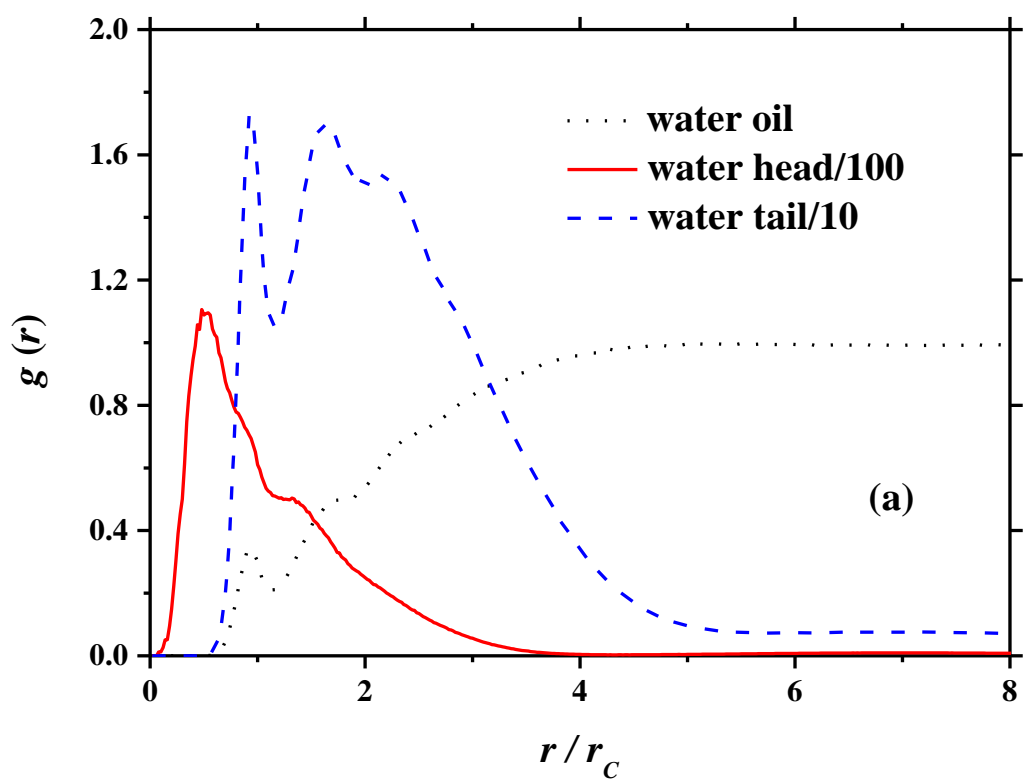

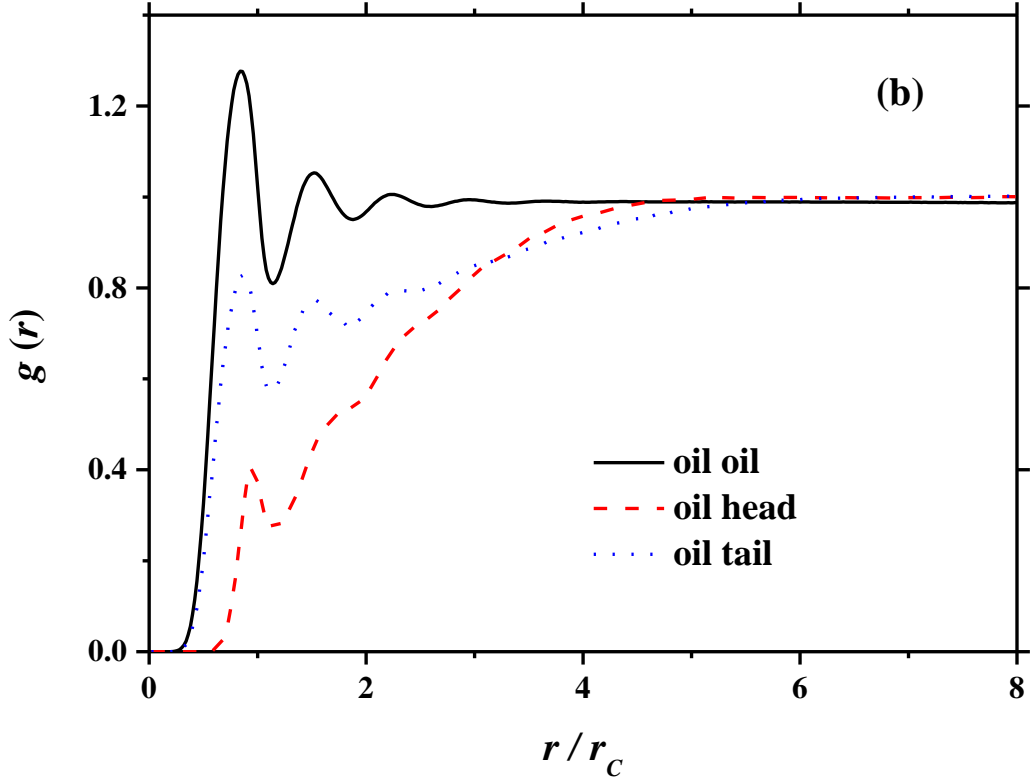

**Figure 5**. (Color online) Radial distribution functions of 100 water beads with 100 ht$_4$ chains in oil. (a) RDFs of water with oil (black, dotted line) and with the ht$_4$ heads (solid red line) and tails (dashed blue line). (b) RDFs of oil with oil (black solid line), and with the heads (red dashed line) and tails (blue dotted line) of ht$_4$. In (a) the RDFs for water − surfactant head, and water − surfactant tail were arbitrarily divided by 100 and 10, respectively, so that they could be plotted on the same scale as the water − oil RDF.

Next, we tracked the number of RMs formed as a function of the number of water molecules in the system. Self − assembly was found in the two nonpolar fluids modeled, oil (Table 1) and scCO$_2$ (Table 2). The effect of different water-surfactant ratios on the process of the aggregate formation was also investigated. Figure 6 plots the ratio $D = N_{aggregates}/N_{surfactants}$ as a function of water-surfactant ratios ($x = N_{water}/N_{surfactants}$) existing during the simulation. For $x = 0$, when water is not present in the system, there is a small quantity of aggregates in oil, while for scCO$_2$ no aggregation is presented. That means that for the case of scCO$_2$ the presence of water is required for aggregate formation [23]. As soon as the number of water beads is increased the formation of aggregates increases also, reaching a plateau, see Fig. 6. After this point, further addition of water beads does not increase the formation of aggregates anymore. Notice that for the case of $ht_4$ in $scCO_2$ the number of surfactant molecules is 50 while in oil the number of $ht_4$ molecules is 100. Less $ht_4$ chains are needed to form aggregates in $scCO_2$ than in oil because $scCO_2$ is a weaker nonpolar solvent.

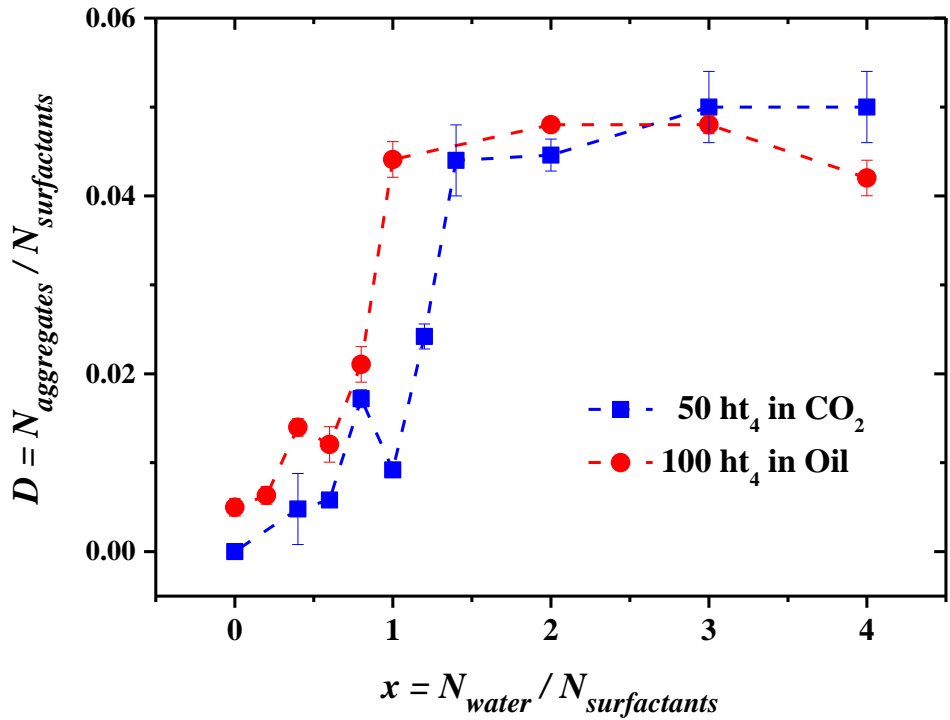

**Figure 6**. (Color online) Number of reverse micelles (aggregates) formed in oil (red circles) and in $scCO_2$ (blue squares) as the number of water beads is increased. The number of surfactant chains is fixed at 50 molecules in $scCO_2$ and 100 molecules in oil.

The comparison between the RDFs of water droplets in oil (red dashed line) and in $scCO_2$ (solid blue line) is shown in Fig. 7. For both solvents, oil and $scCO_2$, there are 50 water beads and 50 $ht_4$ surfactant molecules. Notice the stronger correlation between $scCO_2$ and water than between oil and water, which is the reason why aggregate formation requires more water molecules in $scCO_2$, see Fig. 6.

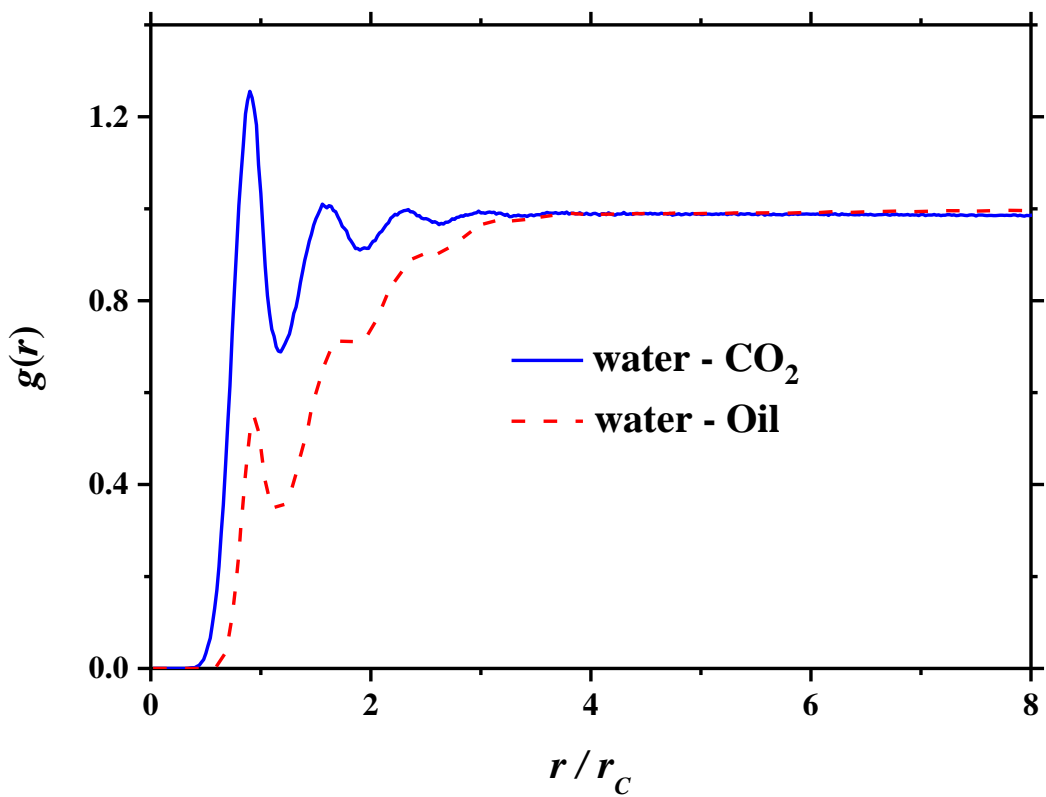

**Figure 7**. (Color online) Radial distribution functions of water beads in oil (red dashed line) and in scCO$_2$ (blue solid line) beads. For these RDFs the number of ht$_4$ chains is fixed at 50 molecules and there are also 50 water beads both in scCO$_2$ and in oil.

Next, the dynamics of aggregation was studied. In Fig. 8 one finds the number of ht$_4$/water aggregates formed in oil and in scCO$_2$ as a function of simulation time. Figure 8 clearly shows the rapid aggregate formation, followed by a quite extended period of stabilization of the aggregates [30]. The aggregates form more rapidly in oil than in scCO$_2$, which is attributed to the difference in the driving forces for aggregation in the two media. The ht$_4$ chains and water bead dissolved in a strong nonpolar solvent (oil) experience significantly stronger aggregation forces than in a less nonpolar solvent (CO$_2$), even at low concentrations. Surfactants in strong nonpolar solvents display faster aggregation dynamics compared with those detected in weak nonpolar or aqueous surfactant systems [9, 30]. Also, higher fluctuations are observed in the scCO$_2$ case. This suggest that it is more difficult to stabilize reverse micelles in scCO$_2$ than in oil [29]. The oscillations in Fig. 8 show that aggregate formation is a dynamic process, with surfactant molecules migrating from one aggregate to another, as well as the water beads, also Fig. 3 and Fig. S1 in the Supplementary Information. This phenomenon corresponds to the multiple equilibrium or open association model of micellization, which proposes a continuous growth of the micelles [61 – 63]. The equilibrium constants can be different in the sequential stages of aggregate formation in nonaqueous solvents, with aggregation numbers being very low at the beginning of the micellization process [64-67].

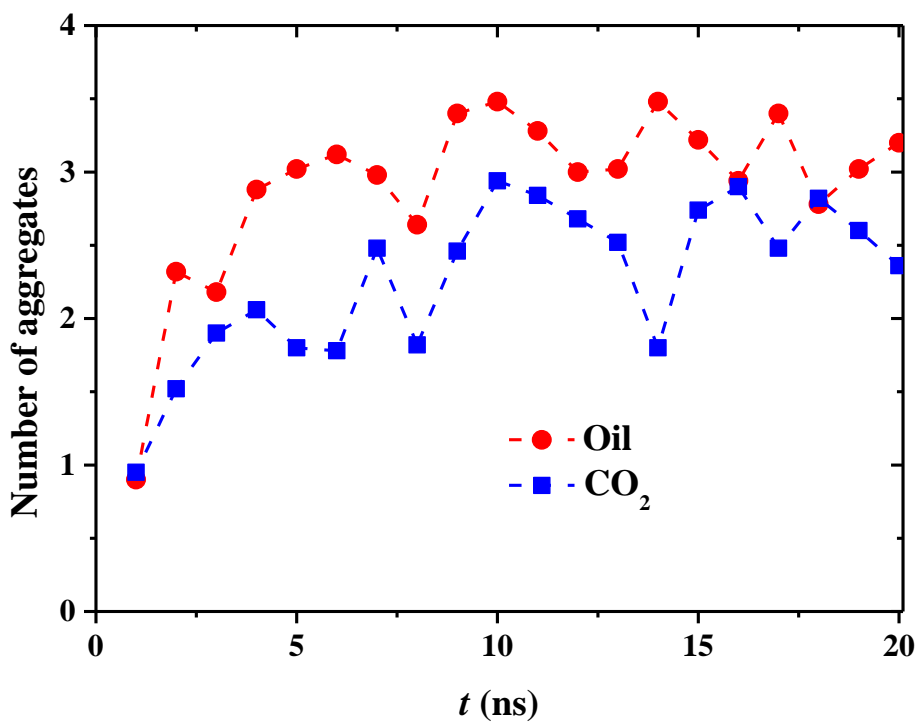

**Figure 8**. (Color online) Number of water/ht$_4$ aggregates formed in oil (red circles) and in CO$_2$ (blue squares) as a function of time. There are 200 water beads and 50 ht$_4$ molecules in both solvents.

The influence of the surfactant architecture on the RM formation was studied as well, increasing the number of beads in the head and the tail of the surfactants. The results can be found in the Supplementary Information. We also investigated the effect of worsening the solvent's quality on the structure of the RM, which is reported in the Supplementary Information.

IV.2 Interfacial tension performance of ht$_4$.

The interfacial tension (IFT) was calculated for an oil/water mix as a function of the concentration of the surfactant ht$_4$. An equal number of water and oil beads was placed in the simulation box, with ht$_4$ molecules added to the system. As the surfactant concentration is increased, the number of water beads and oil beads is reduced equally so that the global number density is always equal to 3. Figure 9 (a) shows a snapshot of this system. Four interfaces between water and oil are formed in the simulation box due to periodic boundary conditions, and surfactants migrate to the interfaces. The surfactants' heads (cyan beads in Fig. 9 (a)) are immersed in the water phase (read beads in Fig. 9 (a)), while the tails (yellow beads in Fig. 9(a)) prefer the oil phase (blue beads in Fig. 9(a)). Density profiles $\rho(z)$ for the system are presented in Fig. 9(b), showing water-oil interfaces (solid red and dashed blue lines in Fig. 9(b) respectively). The density profiles corresponding to the head and tail of the ht$_4$ surfactant along the $z$ – axis of the simulation box are shown in navy dash dotted and dark yellow dash-dot-dotted lines, respectively. There are equal numbers of water beads and oil beads in the system and there are 350 ht$_4$ molecules in the simulation box.

In Fig. 9(b) the lines defining the water – oil interfaces are almost vertical leading to relatively thin interfaces. This is important because it is known that sharp (thin) interfaces produce higher values of the IFT than thicker interfaces [68]. Near the interfaces, there is practically no structuring of water or oil, implying that they are smooth. The hydrophobic beads (tails) of ht$_4$ are accumulated near the oil side of the interfaces, while the hydrophilic beads (heads) agglomerate near the water side of the interface, as is expected.

No surfactant beads are found at the center of the oil or water phases, indicating that all the surfactants molecules migrate to the interfaces.

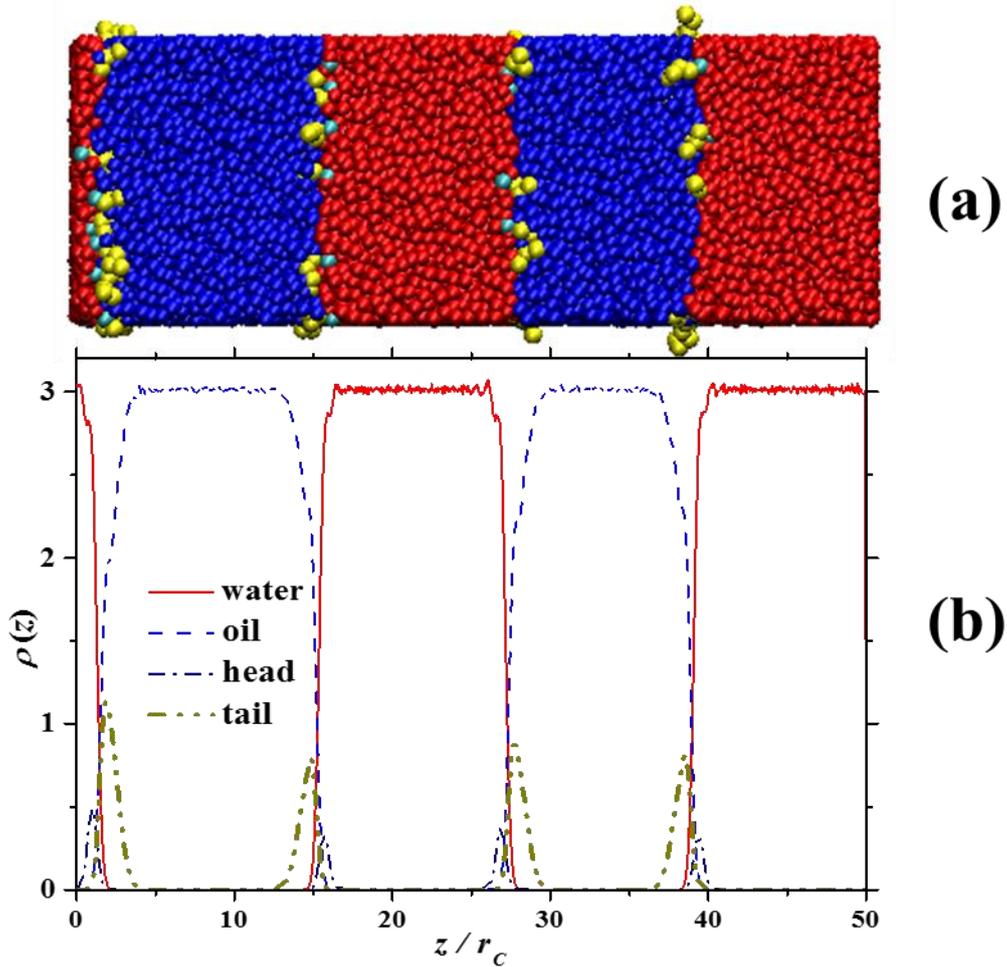

**Figure 9**. (a) (Color online) Snapshot of the oil (blue), water (red) and surfactant (heads in cyan, tails in yellow) beads in the system. There are 350 ht$_4$ surfactant chains and $2.2 \times 10^4$ water beads and $2.2 \times 10^4$ oil beads. (b). (Color online) Density profiles of the components of the system whose snapshot is shown in (a). The system contains water (red solid line), oil (blue dashed line), hydrophilic head (navy dash dotted line) and hydrophobic tail (dark yellow dash-dot-dotted line) beads. There are 350 ht4 chains with $2.2\times10^4$ water beads and $2.2\times10^4$ oil beads.

The IFT of the system was obtained as a function of increasing concentration of the ht$_4$ surfactant. Figure 10 displays the IFT normalized by $\gamma_0$, which is the IFT of the system without surfactants. The ht$_4$ concentration is normalized by the number of molecules of surfactant corresponding with the critical micelle concentration (CMC). The latter is defined as the concentration at which the IFT ceases to decrease upon surfactant addition. The behavior for the IFT as a function of the concentration of surfactant shows a reduction of the IFT which follows a logarithmically decreasing function. This agrees with Szyszkowski's equation [69], as shown in Fig. 10, which assumes that adsorption occurs in monolayers.

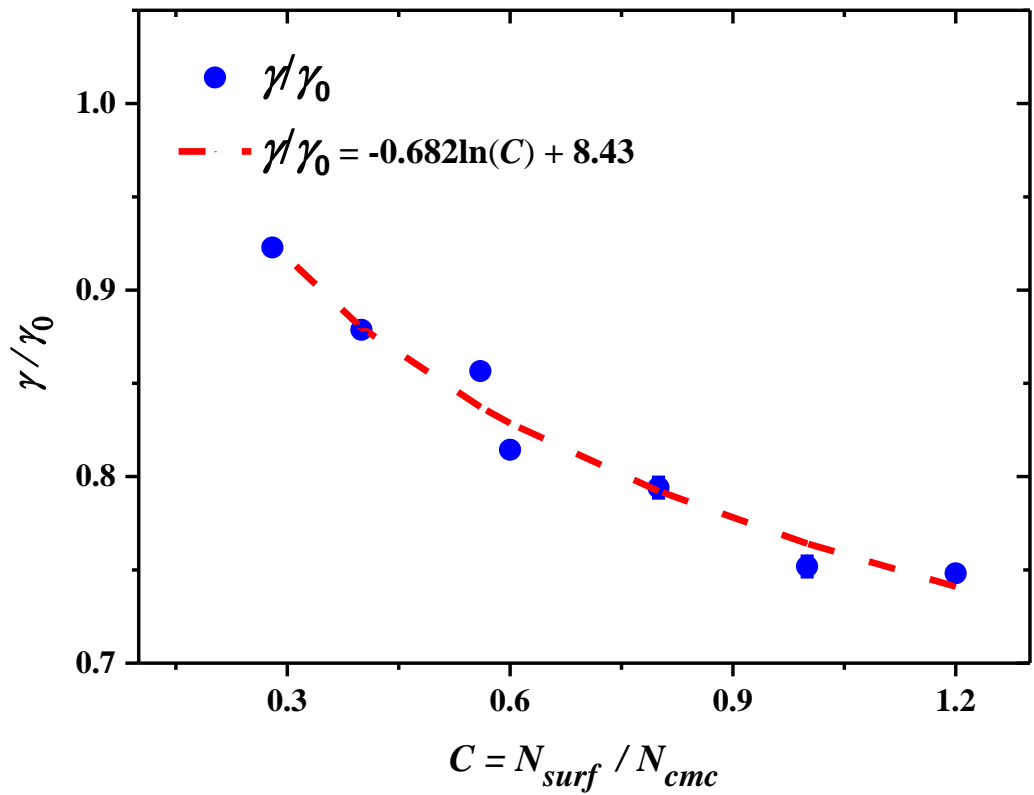

**Figure 10**. (Color online) Oil-water interfacial tension ($\gamma$), normalized by its value when no ht$_4$ surfactants are present ($\gamma_0$), as a function of number of surfactant molecules divided by the number of surfactants at the critical micelle concentration (CMC). The red dashed line shows the fitting of the interfacial tension with Szyszkowski's [78].

Finally, we calculated the potential of mean force ($W_{PMF}$) to obtain further understanding into the structural and thermodynamic properties of the surfactant in the solvents. The $W_{PMF}$ is a many – body, free energy function of the coordinates of two particles with respect to the positions of many other particles [70]. Here it was calculated from the two – dimensional radial distribution functions between pairs of particles, at the water – oil interface. Since the presence of interfaces along the $z-$ direction breaks the three – dimensional symmetry of the system, the radial distribution functions must be evaluated on the $xy-$ plane, where the system is symmetrical due to the periodic boundary conditions. The radial distribution functions were calculated in quasi – two dimensional cylinders of thickness $w = 0.3 r_C$ and radius $r = (1/2)L_x$, where $L_x$ is the length of the simulation box in the $x-$ axis, placed at only one of the water - oil interfaces. The potential of mean force for this system is shown in Fig. 11 and was obtained from the relation $W_{PMF}(r) = -k_B T \ln[g(r/r_C)]$. Here, $k_B$ is Boltzmann's constant and $T$ the absolute temperature. Because the system is modeled with repulsive DPD forces, see Eq. (2), the $W_{PMF}$ curves in Fig. 11 are predominantly repulsive. Nevertheless, there is a small region on the $xy-$ plane of the interface ( $0 \leq r/r_c \leq 1.0$ ) where there is an effective attraction ($W_{PMF} = -0.5 k_B T$) between the tail and head of the surfactants. The strong attraction between the surfactant's head and tail, only at the interface, is the mechanism that makes this surfactant effective. That is because this effective attraction (which arises when there are water – oil interfaces) between the surfactant's head and tails drives them all to the interfaces (see Fig. 9(b)), rather than to forming micelles. This process reduces the IFT.

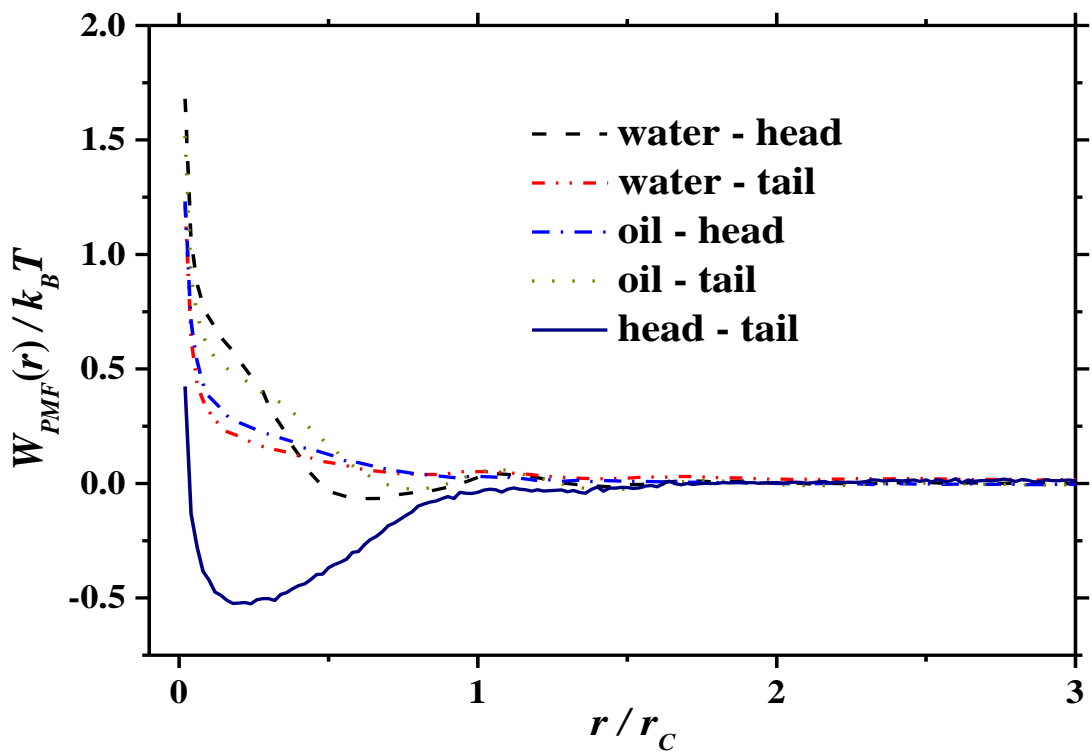

**Figure 11**. (Color online) Two – dimensional potentials of mean force ($W_{PMF}/k_BT$) for all the components in the system, calculated at water/oil interface.

The attractive well in the potential of mean force between the surfactants' heads and tails (see the solid dark blue line in Fig. 11) is due to the many – body interactions between all the head and tail beads at the water – oil interface. It is not due to the bonding interaction between the head and tail of a surfactant molecule because the conservative DPD repulsive force between the head and tail beads is larger than the harmonic attraction between them. The attractive well between head and tail beads in the $W_{PMF}$ emerges from their close spatial proximity at the oil/water interface. This, in turn, occurs because of the high affinity of the head beads for water and the tail beads for oil.

## V CONCLUSIONS

In this work we report the self – association properties of a model linear surfactant under two situations: dissolved in solvents (oil and scCO$_2$), and at the oil – water interface, using mesoscopic simulations. Small amounts of water were found to be favorable for the self-assembly of the surfactants into reverse micellar structures. The self-assembly of ht$_4$ in both solvents was driven mainly by two factors. The strong correlation of water with the surfactant's heads and, at the same time, by the strong attraction of the nonpolar solvent with the surfactant's tails. The rate of micellization was greater when the surfactants are dissolved in oil than in scCO$_2$, when small quantities of water are added. This is attributed to the difference in the driving forces for aggregation in two media. The surfactant and water molecules dissolved in a strong nonpolar solvent (oil) undergo stronger aggregation than in a less nonpolar solvent (scCO$_2$), even at low concentrations. Surfactants in strong nonpolar solvents display faster aggregation dynamics than those found in weak nonpolar or aqueous systems. Fluctuations in aggregate formation observed in the scCO$_2$ case suggest that it is more difficult to form reverse micelles in scCO$_2$ than in oil because of the affinity between water molecules and CO$_2$. Such affinity is unfavorable for water molecules grouping in the nucleus of aggregates, which is necessary for the reverse micellization process. The IFT shows a reduction as a function of surfactant concentration, which follows a logarithmically decreasing function. This result agrees with Szyszkowski's equation at low surfactant concentrations which assumes that adsorption occurs in monolayers. The $W_{PFM}$ shows that the effective attraction between the surfactants' heads and tails, only at the interface, is the mechanism that makes this surfactant effective in reducing the IFT.

## CONFLICTS OF INTEREST

There are no conflicts of interest to declare.

## ACKNOWLEDGEMENTS

We acknowledge funding from the European Union's Horizon 2020 Program under the ENERXICO Project, grant agreement No. 828947 and under CONACYT-SENER-Hidrocarburos (Mexico), grant agreement No. B-S-69926. The calculations reported here were mostly performed using the supercomputing facilities of ABACUS Laboratorio de Matemática Aplicada y Cómputo de Alto Rendimiento of CINVESTAV-IPN. The authors also acknowledge the computer resources offered by the Laboratorio de Supercómputo y Visualización en Paralelo (LSVP − Yoltla) of UAM − Iztapalapa, where some of the simulations were carried out.